\documentclass[twoside]{article}
\usepackage{graphicx}
\usepackage{fancyhdr}
\usepackage{multicol}
\usepackage{styl-ap}

\begin{document}
\title{Opera Highlights}
\author{Thomas Strauss\work{1}, for the OPERA collaboration}
\workplace{Albert Einstein Center for Fundamental Physics, Laboratory for High Energy Physics, University of Bern, Sidlerstrasse 5, CH-3012 Bern
}
\mainauthor{thomas.strauss@lhep.unibe.ch}
\maketitle

\begin{abstract}
The OPERA experiment is a long baseline neutrino oscillation experiment aimed at observing the $\nu_{\mu} \rightarrow \nu_{\tau} $ neutrino oscillation in the CERN neutrino to Gran Sasso beamline in the appearance mode by detecting the $\tau$ - decay. Here I will summarize the results from the run years 2008-10 with an update on observed rare decay topologies and the results of the neutrino velocity measurements.
\end{abstract}


\begin{multicols}{2}
\section{Introduction}
The OPERA experiment \cite{1}, located at the Gran Sasso laboratory (LNGS), aims at observing the $\nu_{\mu}\rightarrow\nu_{\tau}$ neutrino oscillation in the direct appearance mode in the CERN neutrino to Gran Sasso (CNGS) \cite{3a,3b} beam by detecting the decay of the $\tau$ produced in charged current (CC) interactions. A detailed description of the detector can be found in \cite{1,2a,2b,2c,2d,2e,2f}. The OPERA detector consists of two identical  Super Modules (SM), each of them consisting of a target area and a muon spectrometer, as shown in Fig. \ref{strauss-fig1}. The target area consists of alternating layers of scintillator strip planes and target walls. The muon spectrometer is used to reconstruct and identify muons from $\nu_{\mu}$-CC interactions and estimate their momentum and charge. 

\begin{myfigure}
\centerline{\resizebox{70mm}{!}{\includegraphics{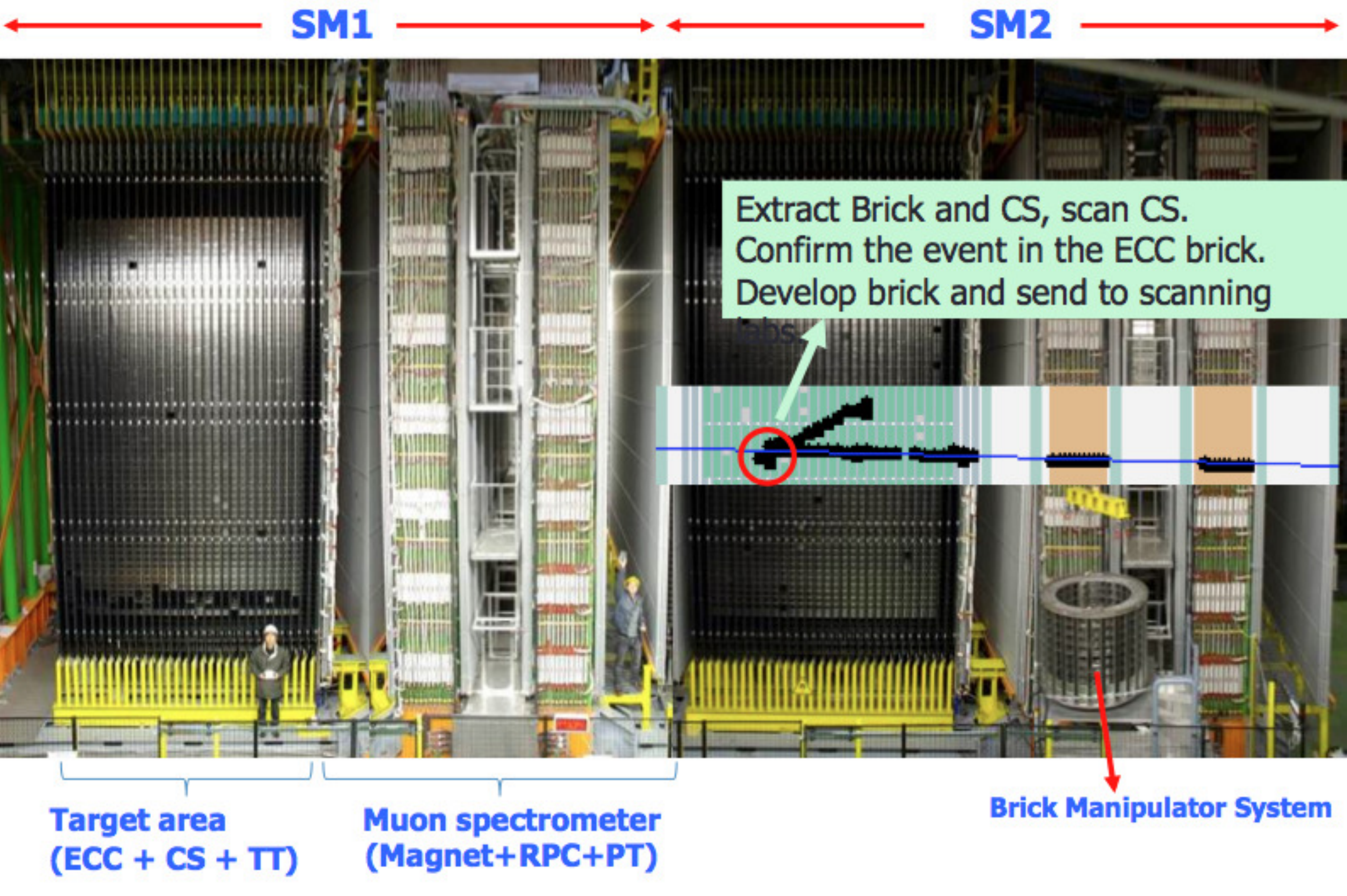}}}
\caption{Picture of the OPERA detector, with a view of a reconstructed neutrino interaction occurring in the 2nd Super Module.}
\label{strauss-fig1}
\end{myfigure}

\begin{myfigure}
\centerline{\resizebox{70mm}{!}{\includegraphics{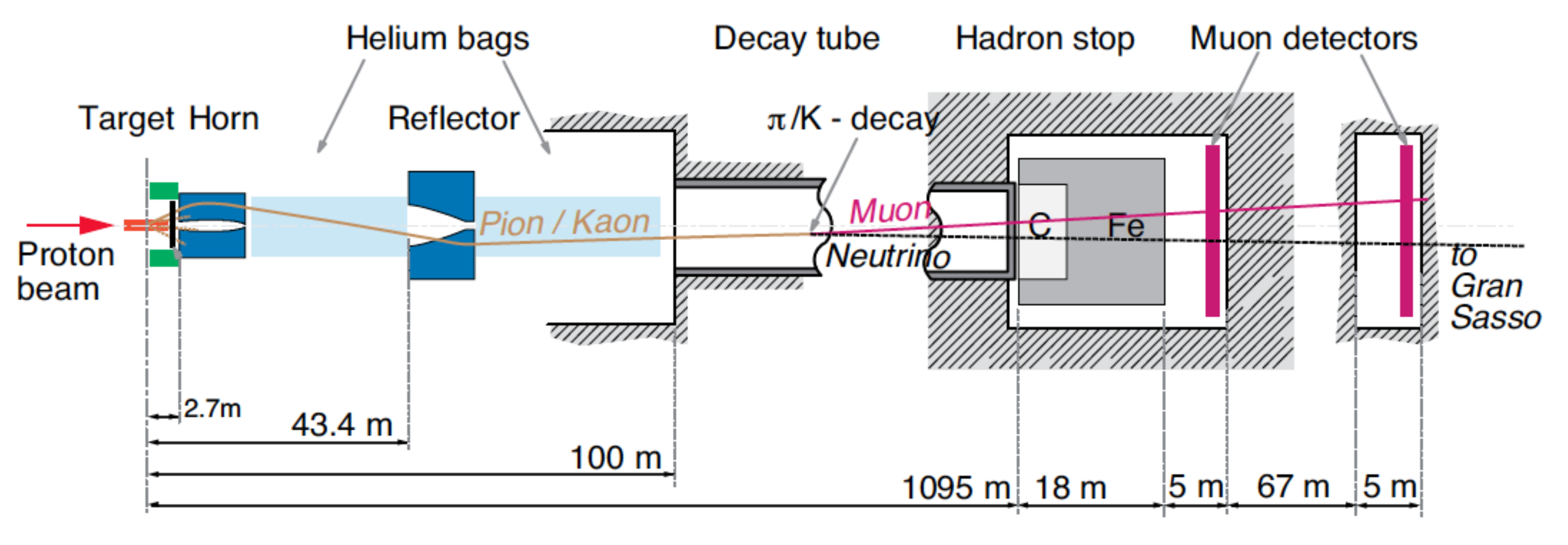}}}
\caption{The CNGS neutrino beamline. Figure from \cite{6}.}
\label{author-fig4}
\end{myfigure}

The target walls are trays in which target units of $10\times12.5\mathrm{ cm^2}$ and a depth of 10\,$X_0$ in lead (7.9\,cm)	 with a mass of around 10~kg each are stored: they are also refered to as bricks. A brick is formed by alternating layers of lead plates and emulsion films (2 emulsion layers separated by a plastic base) building an emulsion cloud chamber (ECC). This provides high granularity and high mass, which is ideal for $\nu_{\tau}$ interaction detection. Fig. \ref{strauss-fig3} left shows an image of the unwrapped ECC brick. On the right, the arrangement of the scintillator strip planes (Target Tracker, TT) and the ECC is shown. Note an extra pair of emulsion films in a removable box called a changeable sheet (CS), shown in blue. The total mass of each target area is about 625\,tons, leading to a target mass of 1.25\,ktons for 145'000 bricks.

\begin{myfigure}
\centerline{\resizebox{70mm}{!}{\includegraphics{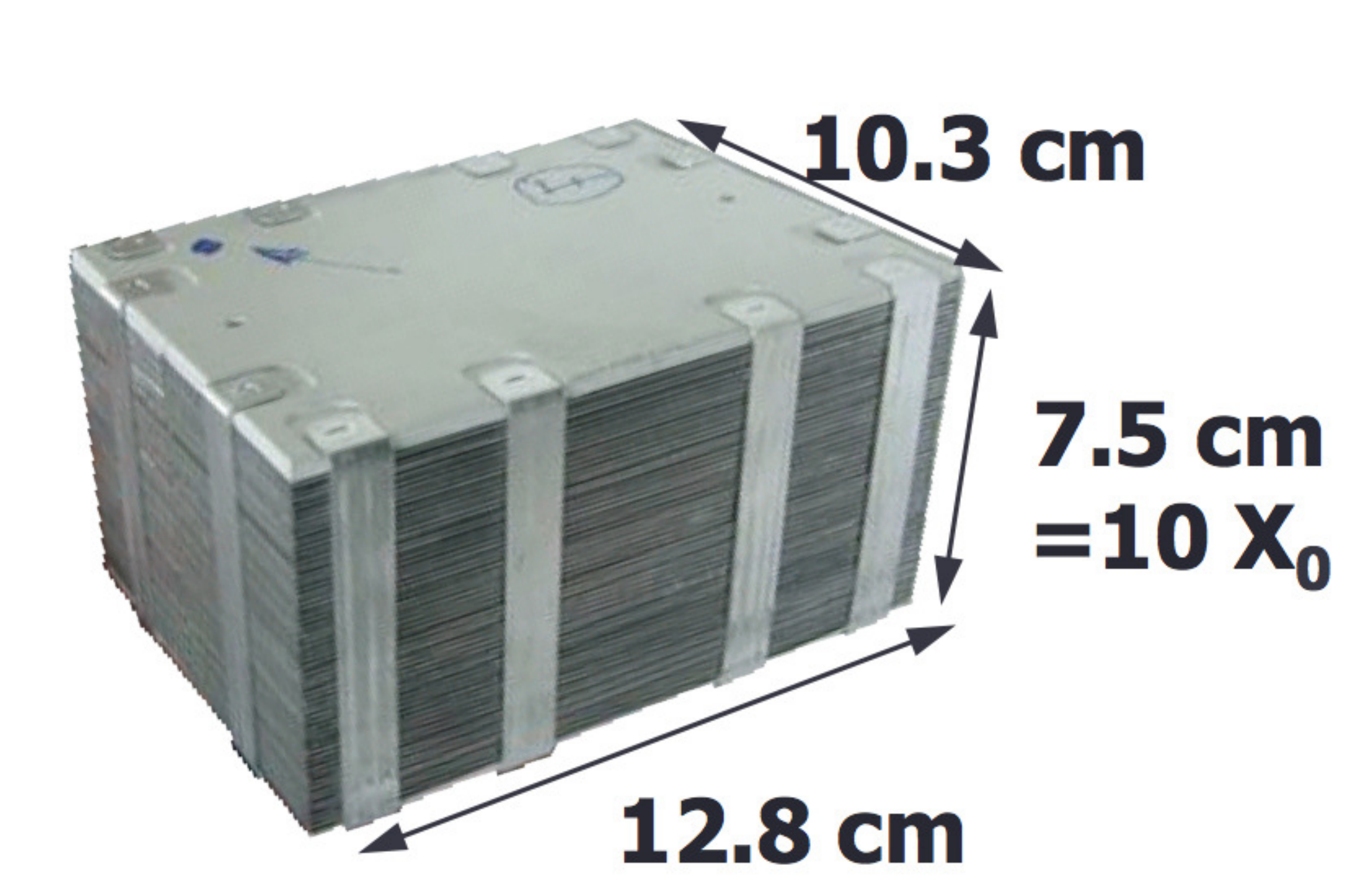}\includegraphics{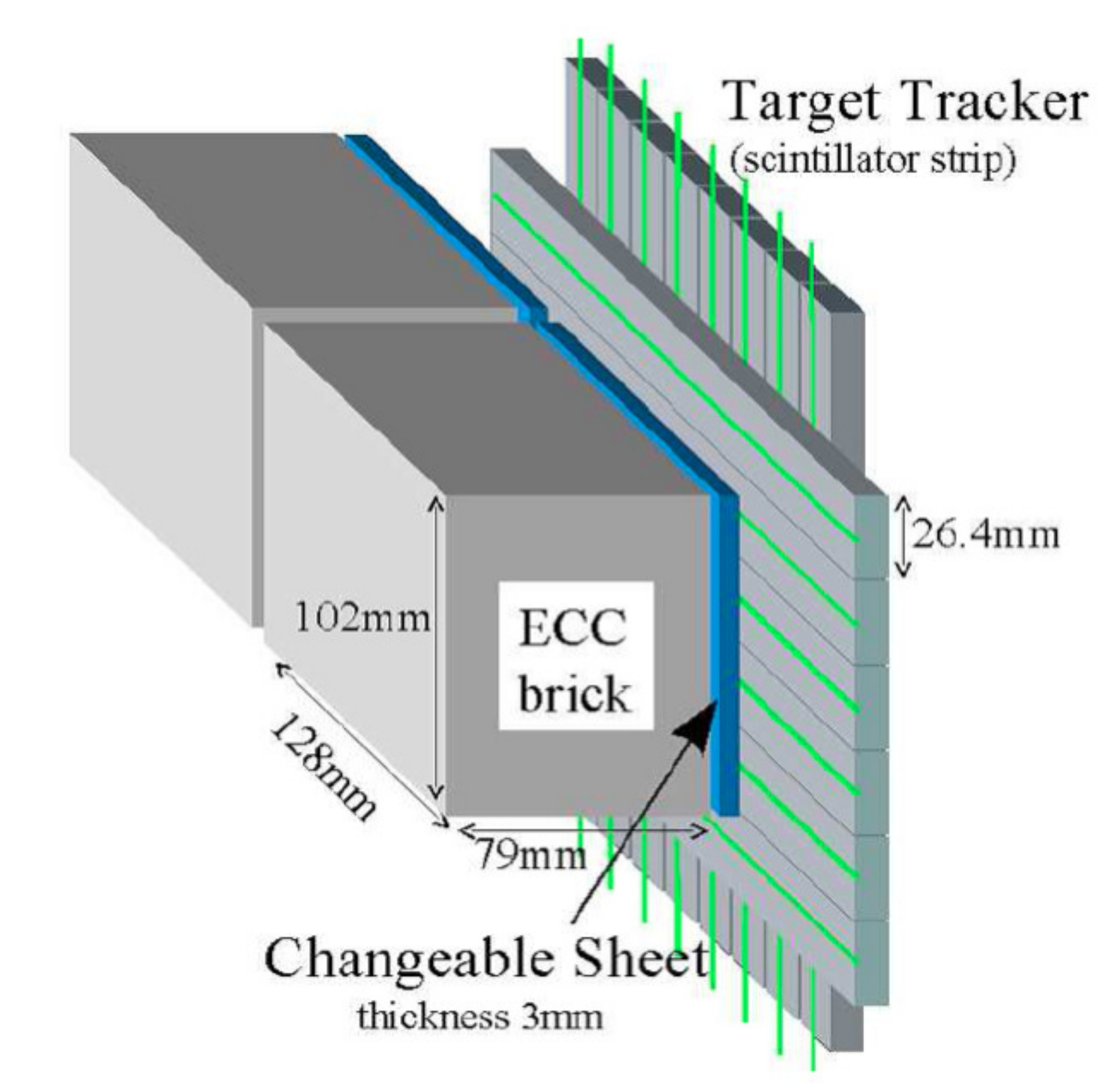}}}
\caption{$\tau$ detection principle in OPERA.}
\label{strauss-fig3}
\end{myfigure}

\section{DAQ and analysis}
The information from the reconstructed event, recorded by the electronic detectors, is used to predict the most probable ECC for the neutrino interaction vertex \cite{4}. A display of a reconstructed event is indicated in Fig. \ref{strauss-fig1}. Fig. \ref{strauss-fig2} shows the procedure for localizing the event vertex in the ECC, by extrapolating the reconstructed tracks from the electronic detector to the CS emulsion films. 

\begin{myfigure}
\centerline{\resizebox{70mm}{!}{\includegraphics{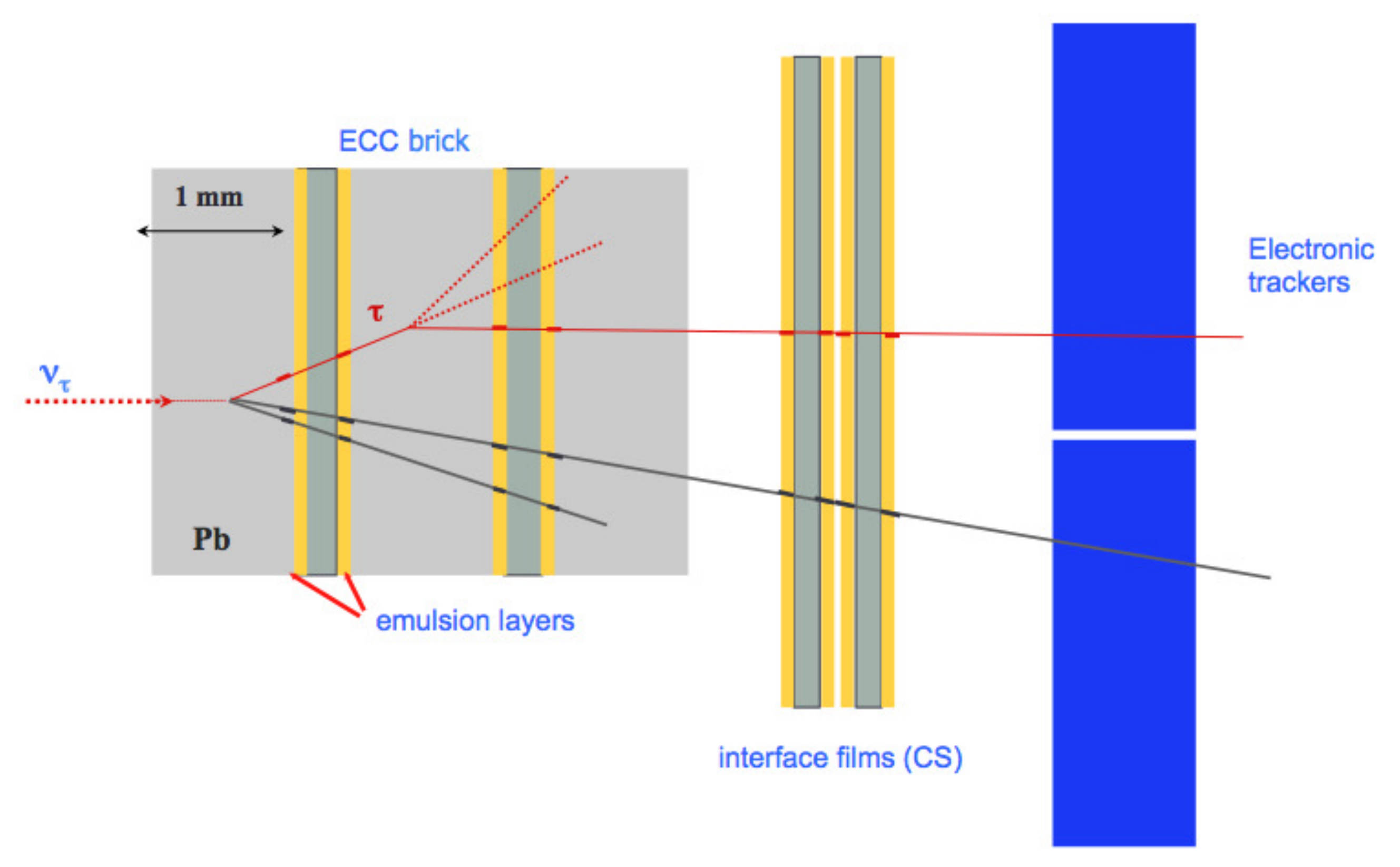}}}
\caption{Event detection principle in the OPERA experiment.  The candidate interaction brick is determined from the prediction of the electronic detector (blue). The changeable sheet (CS) is used to confirm the prediction. From the CS result, the tracks are followed up to the interaction point inside the ECC.}
\label{strauss-fig2}
\end{myfigure}

The signal recorded in the CS films will confirm the prediction from the electronic detector, or will act as a veto and trigger a search in neighboring bricks to find the correct ECC in which the neutrino interaction is contained. After a positive CS result, the ECC will be unpacked and the emulsion films are developed and sent to one of the various scanning stations in Japan and Europe. Dedicated automatic scanning systems allow us to follow the tracks from the CS prediction up to their stopping point. Around these stopping points, a volume of 1\,cm$^2$ times 15\,emulsion films will be scanned to find the interaction vertex. As illustrated in Fig. \ref{strauss-fig2}, only track segments in the active emulsion volume are visible and a reconstruction of the event is needed to find tracks and vertices. A dedicated procedure, called a ``decay search" is used to search for possibly interesting topologies, like the $\tau$-decay pictured in Fig. \ref{strauss-fig2}. The accuracy of the track reconstruction goes from cm in the electronic detector, down to mm for the CS analysis and to micrometric precision in the final vertex reconstruction (after aligning the ECC emulsion plates with passing-through cosmic-ray tracks).
\newline
\subsection{Tau detection} $\tau$ detection is only possible due to the micrometric resolution of the emulsion films, as it allows us to separate the primary neutrino interaction vertex from the decay vertex of the $\tau$ particle. The most prominent background for $\tau$ decay is either hadron scattering or charged charm decays. The background from hadron scattering can be controlled by cuts applied to the event kinematics. The background due to charm can be reduced by identifying the muon at the primary vertex, as charm will occur primarily in $\nu_{\mu}$-CC interactions (for further details see \cite{2e,4}). After topological and kinematical cuts are applied, the number of background events in the nominal events sample is anticipated to be 0.7 at the end of the experiment.

In 2010 the first $\nu_{\tau}$ candidate was reconstructed inside the OPERA emulsion. It was recorded in an event classified as a neutral current, as no muon was identified in the electronic detector. To crosscheck the $\tau$ hypothesis, all tracks were followed downstream of the vertex until their stopping or their re-interaction point. They were all attributed to be hadrons, and no soft muon ($E<2$\,GeV) was found. In 2010, the expected total number of $\tau$ candidates was 0.9, while the expected background was less than 0.1 events. More details on the analysis are presented in \cite{2e}. In Fig. \ref{strauss-fig8} shows a picture of this event. Track number 4, labeled as the parent, is the $\tau$, decaying into one charged daughter track. The two showers are most likely connected to the decay vertex of the $\tau$, rather than activity from the primary interaction. Thus the decay is compatible with: $\tau\rightarrow\eta^-(\pi^-+\pi_0)\nu_{\tau}$. 

\begin{myfigure}
\centerline{\resizebox{70mm}{!}{\includegraphics{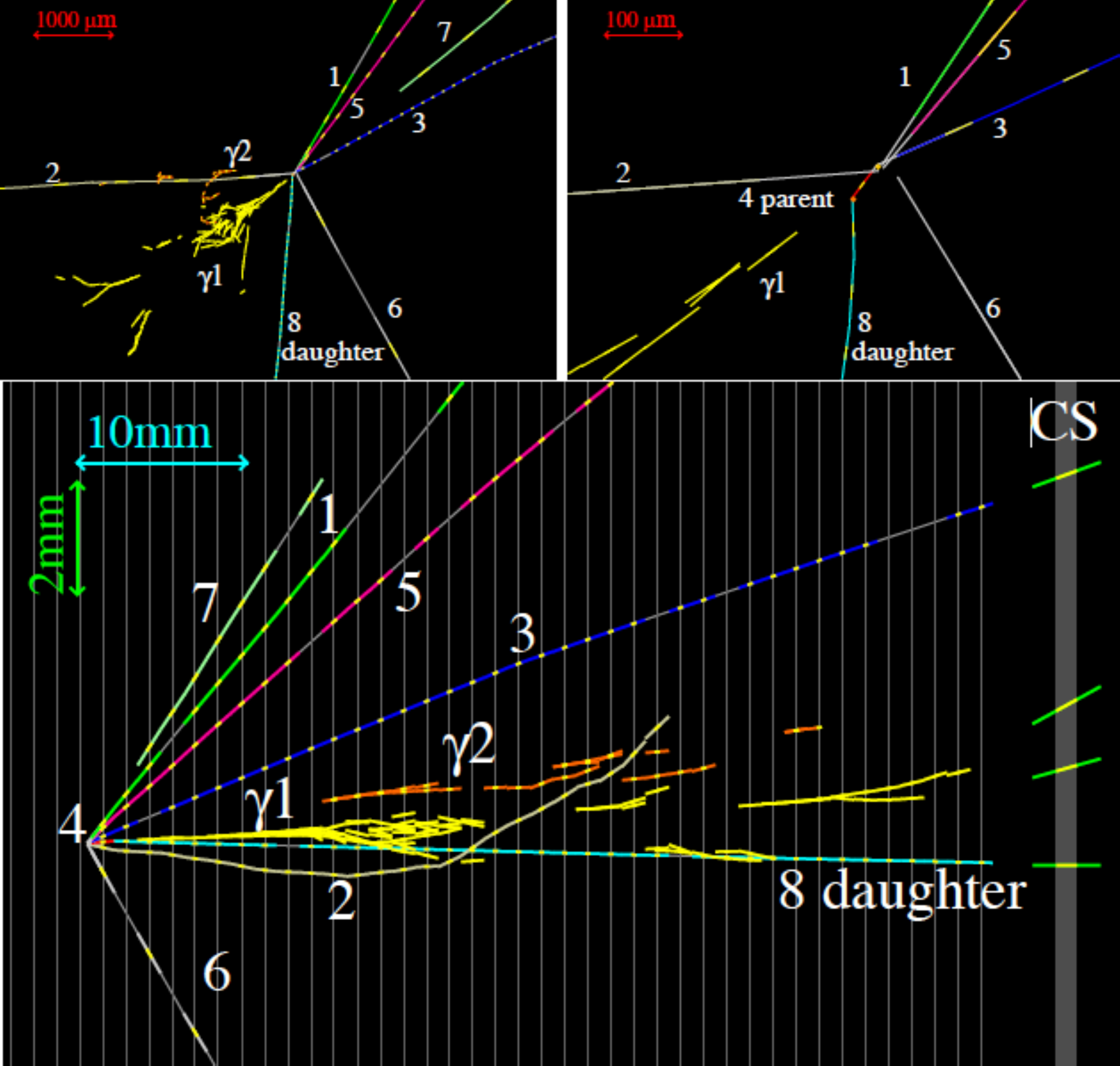}}}
\caption{Display of the 2010 $\tau^-$ candidate event. Top left: view transverse to the neutrino
direction. Top right: same view zoomed on the vertices. Bottom: longitudinal view. Figure from \cite{2e}.}
\label{strauss-fig8}
\end{myfigure}

\begin{myfigure}
\centerline{\resizebox{70mm}{!}{\includegraphics{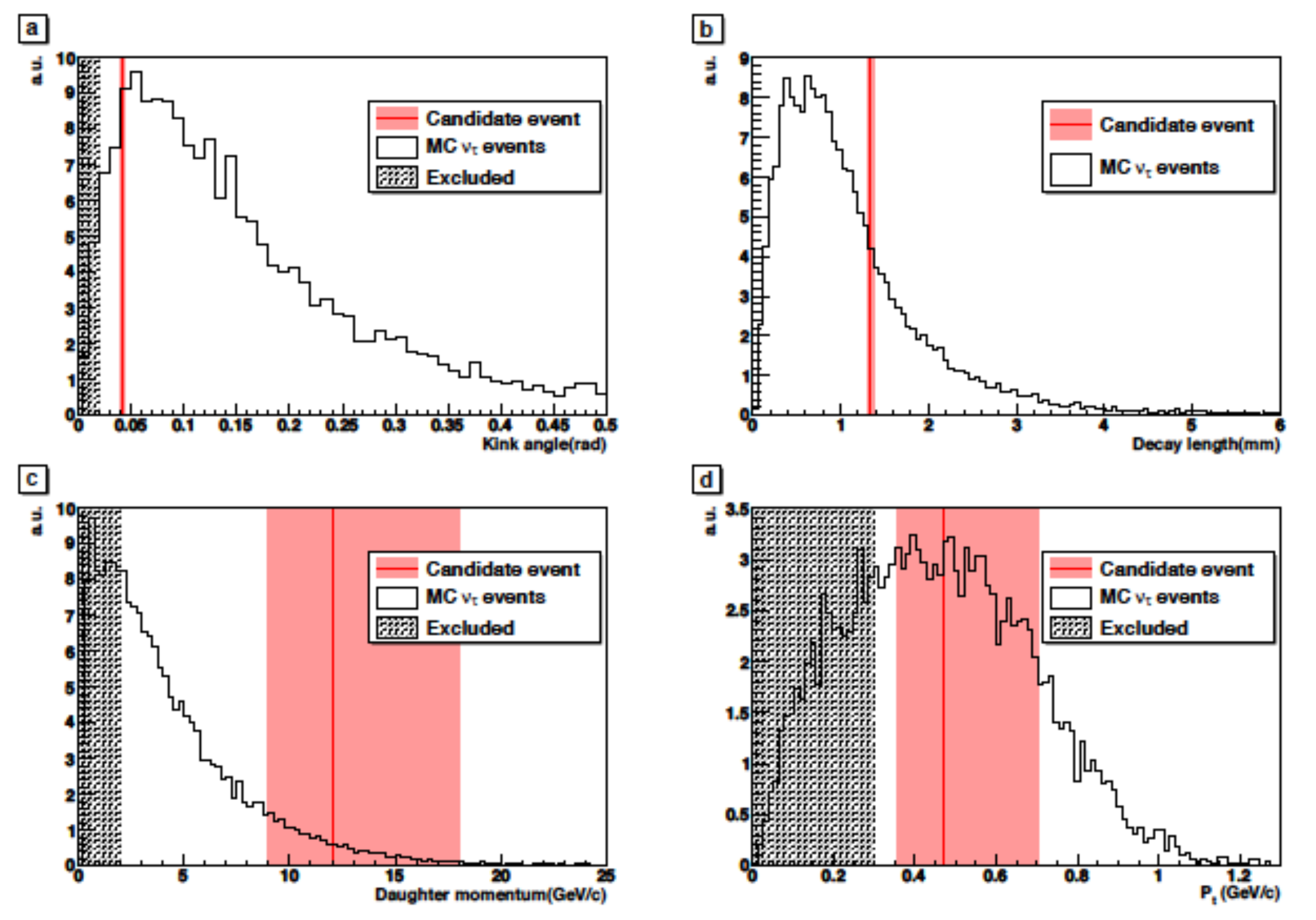}}}
\caption{MC distribution of: a) the kink angle for $\tau$ decays, b) the path length of the $\tau$, c) the momentum of the decay daughter, d) the total transverse momentum $P_T$ of the detected daughter particles of $\tau$ decays with respect to the parent track. The red band shows the 68.3\,\% of values allowed for the candidate event, and the dark red line indicates the
most probable value. The dark shaded area represents the excluded region corresponding to the a priori $\tau$ selection cuts. Figure from \cite{2e}.}
\label{strauss-fig9}
\end{myfigure}

Fig. \ref{strauss-fig9} shows the cuts used for the selection criteria defined at the time of the proposal and the kinematic variables of the $\tau$ decay observed by the OPERA experiment. At the time of this conference, the number of expected $\tau$ candidates was 1.7, with 0.5 events expected in the single-prong channel. The expected background for the analyzed event sample corresponding to $4.9\times10^{19}$\,protons on target (pot) was $0.16\pm0.05$ events. 

At the time of writing these proceedings, a second $\tau$ candidate appeared \cite{7a}.

\subsection{Physics run performance and data analysis status}
Since 2007, the OPERA experiment has collected a total of $18.5\times10^{19}$\,pot. This corresponds to about 15'000 interactions in the target areas of the experiment. Fig. \ref{strauss-fig6} shows from top to bottom, as a function of time, the integrated number of events occurring in the target (showing the CNGS shutdown periods),  with their vertex reconstructed by the electronic detectors, for which at least one brick has been extracted, for which at least one CS has been analysed, for which this analysis has been positive (track stubs corresponding to the event have been found), for which the brick has been analysed, for which the vertex has been located, for which the decay search has been completed.

The efficiency of the analysis of the most probable CS is rather low, about 65\,\%, and is significantly lower for NC events, among which $\nu_{\tau}$ interactions are the most likely to be found. To recover this loss, multiple brick extraction is performed; this brings the final efficiency of observing tracks of the event in the CS to about 74\,\%. The efficiency for locating an event seen on the CS in the brick is about 70\,\%. At the time of this conference, a total of 4611 events has been localised in the bricks and the decay search has been completed for 4126 of them.

\begin{myfigure}
\centerline{\resizebox{70mm}{!}{\includegraphics{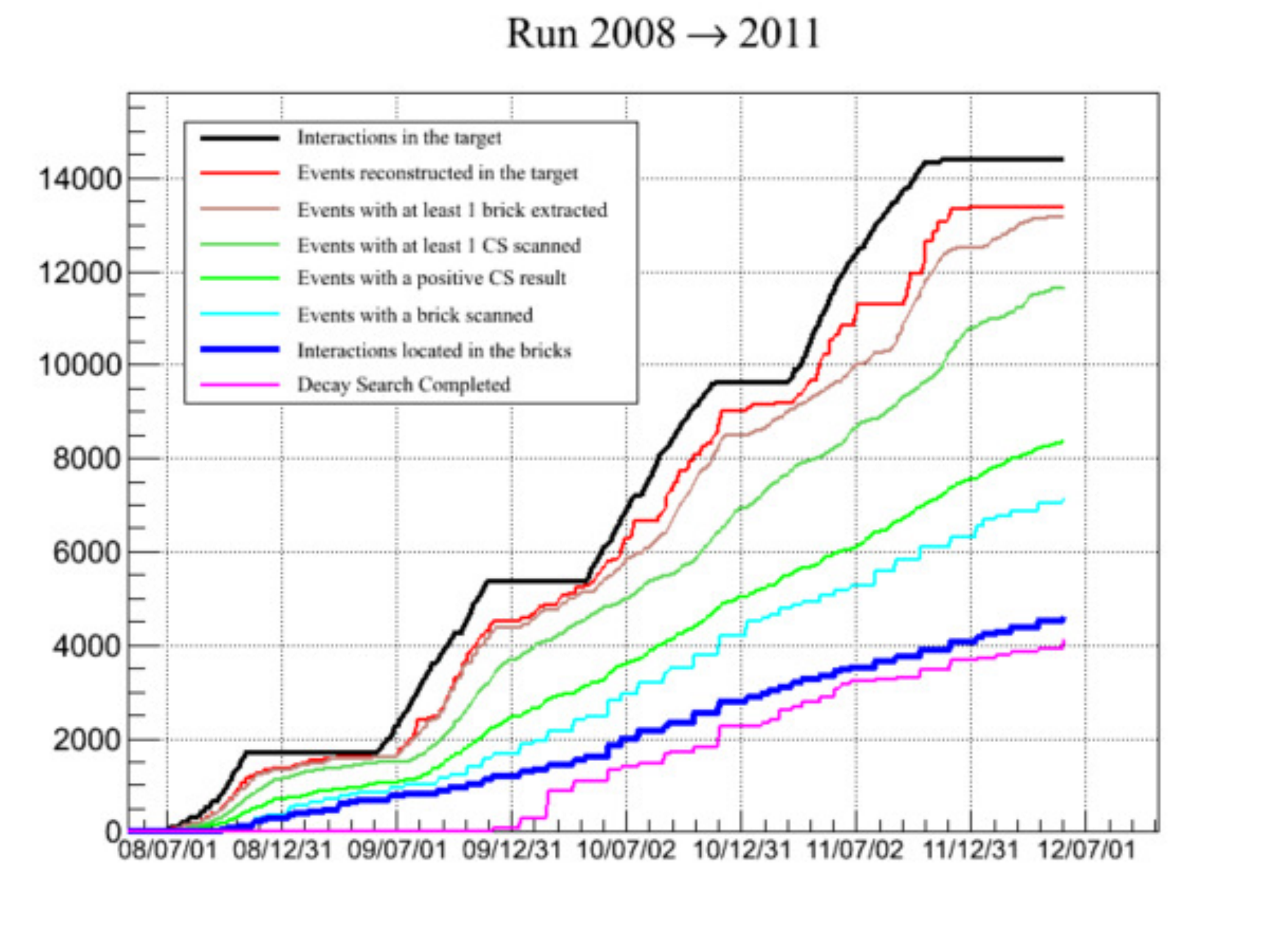}}}
\caption{Events recorded and analyzed in the OPERA experiment from 2008 until 2012.}
\label{strauss-fig6}
\end{myfigure}
After an ECC has been identified to be the most likely interaction brick, the ECC is developed and sent to one of the scanning laboratories, where it is scanned within a short time. The efficiency for locating an event within the ECC is about 70\,\% with respect to the number of positive CS results. After the event has been located, a dedicated decay search is performed to obtain a data sample which can be compared to MC and which provides uniform data quality from all laboratories. This decay search includes the search for decay daughters and a reconstruction of the kinematics of the particles at the vertex. At the time of this conference, 4611 events have been localized in the ECC, with a total of 4126 CC and NC events having completed the decay search.

\subsection{Charm decay topologies in neutrino interactions}
In about 5\,\% of the $\nu_{\mu}$-CC interactions, the production of a charmed particle at the primary vertex takes place. Charmed particles have lifetimes similar to the $\tau$ and similar decay channels. Thus charm events provide a subsample of decay topologies similar to $\tau$ decay, for which the detection efficiency can be estimated based on MC simulation. A study of a high purity selection of charm events in  2008 and 2009 shows agreement with the data \cite{5}. One-prong charm decay candidates are retained if the charged daughter particle has a momentum larger than 1 GeV/c. This leads to an efficiency of $\epsilon_{\mbox{short}} = 0.31 \pm 0.02 (\mbox{stat.})\pm 0.03 (\mbox{syst.})$ for short and $\epsilon_{\mbox{long}} = 0.61 \pm 0.05 (\mbox{stat.}) \pm 0.06 (\mbox{syst.})$ for long charm decays, wherein `long' means the production and the decay vertices are not located in the same lead plate. The number of events for which the charm search is complete is 2167 CC interactions. In these we expect $51\pm7.5$ charm candidates, with a background of $5.3\pm2.3$ events. The number of observed candidates is 49, which is in agreement with expectations. 

\begin{myfigure}
\centerline{\resizebox{70mm}{!}{\includegraphics{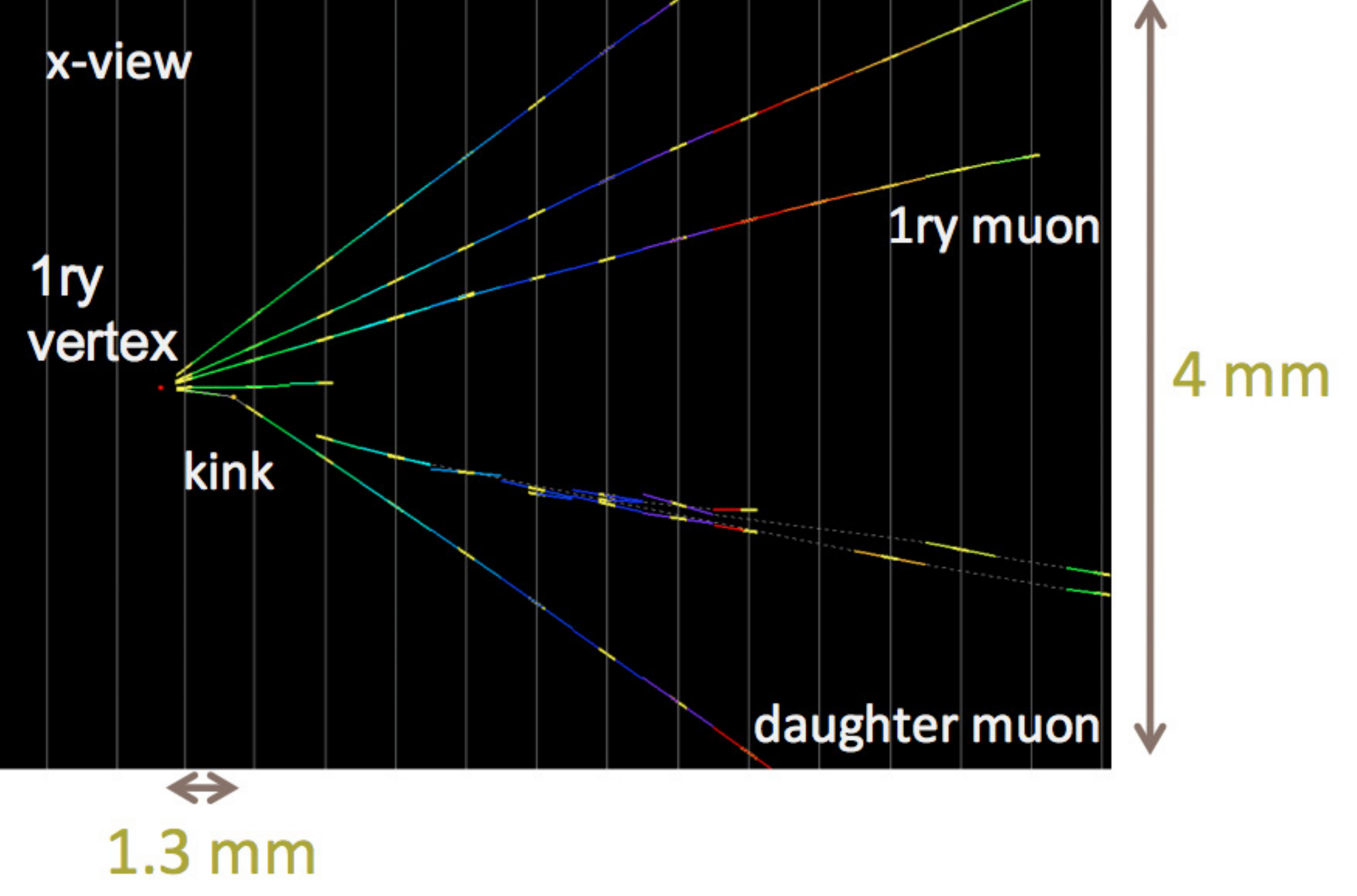}}}
\centerline{\resizebox{70mm}{!}{\includegraphics{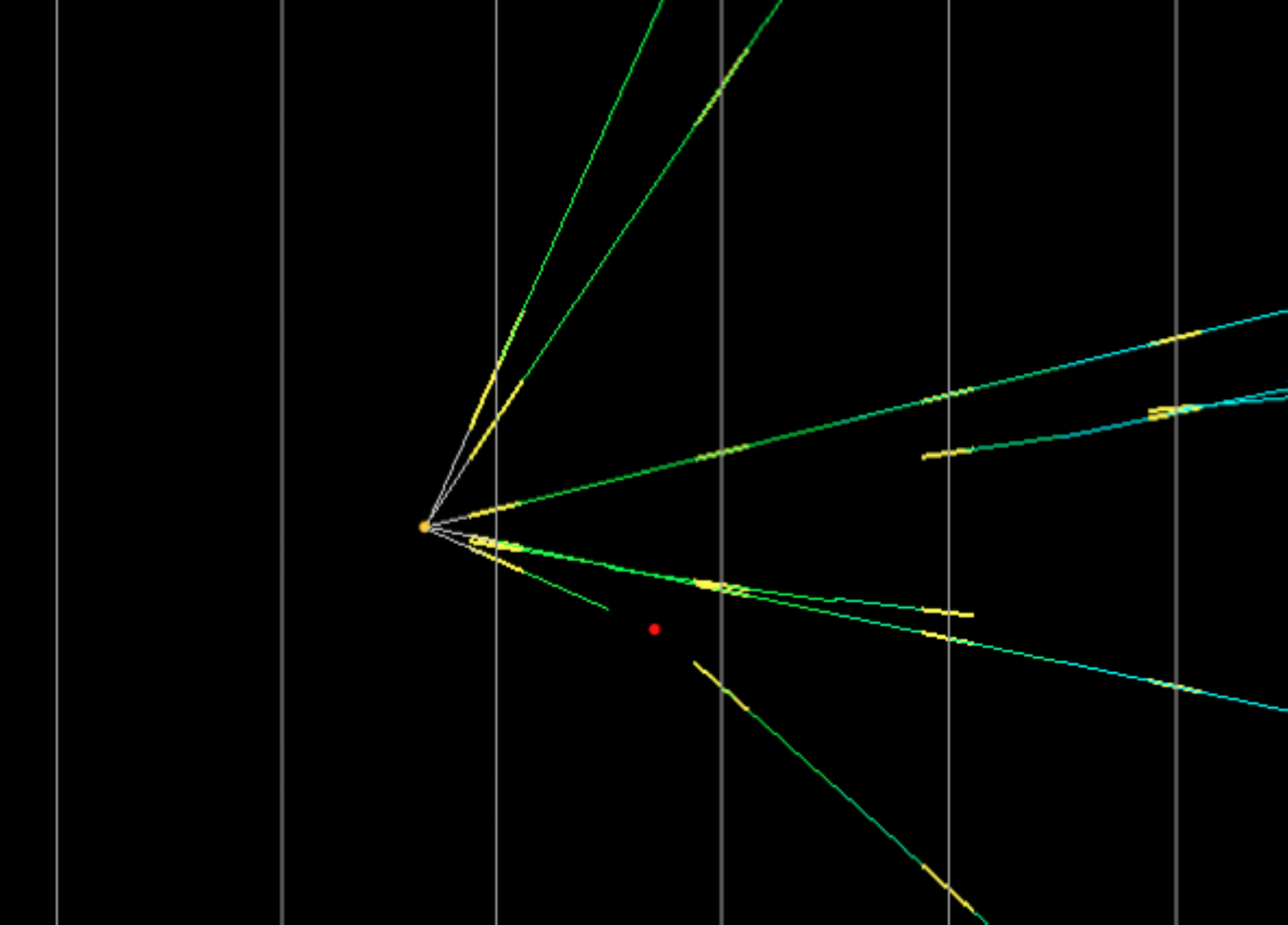}}}
\caption{Display of one of the charm events. Top: View of the reconstructed event in the emulsion. Bottom: Zoom in the vertex region of the primary and secondary vertex. }
\label{strauss-fig10}
\end{myfigure}

Fig. \ref{strauss-fig10} shows a charm decay detected in the OPERA experiment. In the electronic detector reconstruction, two muons were observed, one charged positively, the other negatively. The $\mu^-$ is attached to the primary vertex, while the $\mu^+$ is connected to the decay vertex. This topology corresponds with a charged charm decaying into a muon and the measured kinematic parameters are a flight length of 1330\,$\mu$m and a kink angle of 209\,mrad. The impact parameter (IP) of the $\mu^+$ with respect to the primary vertex is 262\,$\mu$m, and its momentum is measured as 2.2\,GeV/c. This accounts for a transverse momentum ($P_T$) of 0.46\,GeV/c.

\section{$\nu$-velocity measurement}
Due to the time structure of the CERN SPS beam, the OPERA experiment is able to trigger on the proton spill hitting the CNGS target. As a result, the electronic detector provides a time signal of the recorded events, which can be used to measure the neutrino velocity in the CNGS beam. One needs to measure with precision of some ns the flight time (time of flight - TOF) between CERN and LNGS, and the distance between reference points in the detector and the CNGS. The concept of the neutrino time of flight measurement is illustrated in Fig. \ref{strauss-fig15}.

\begin{myfigure}
\centerline{\resizebox{70mm}{!}{\includegraphics{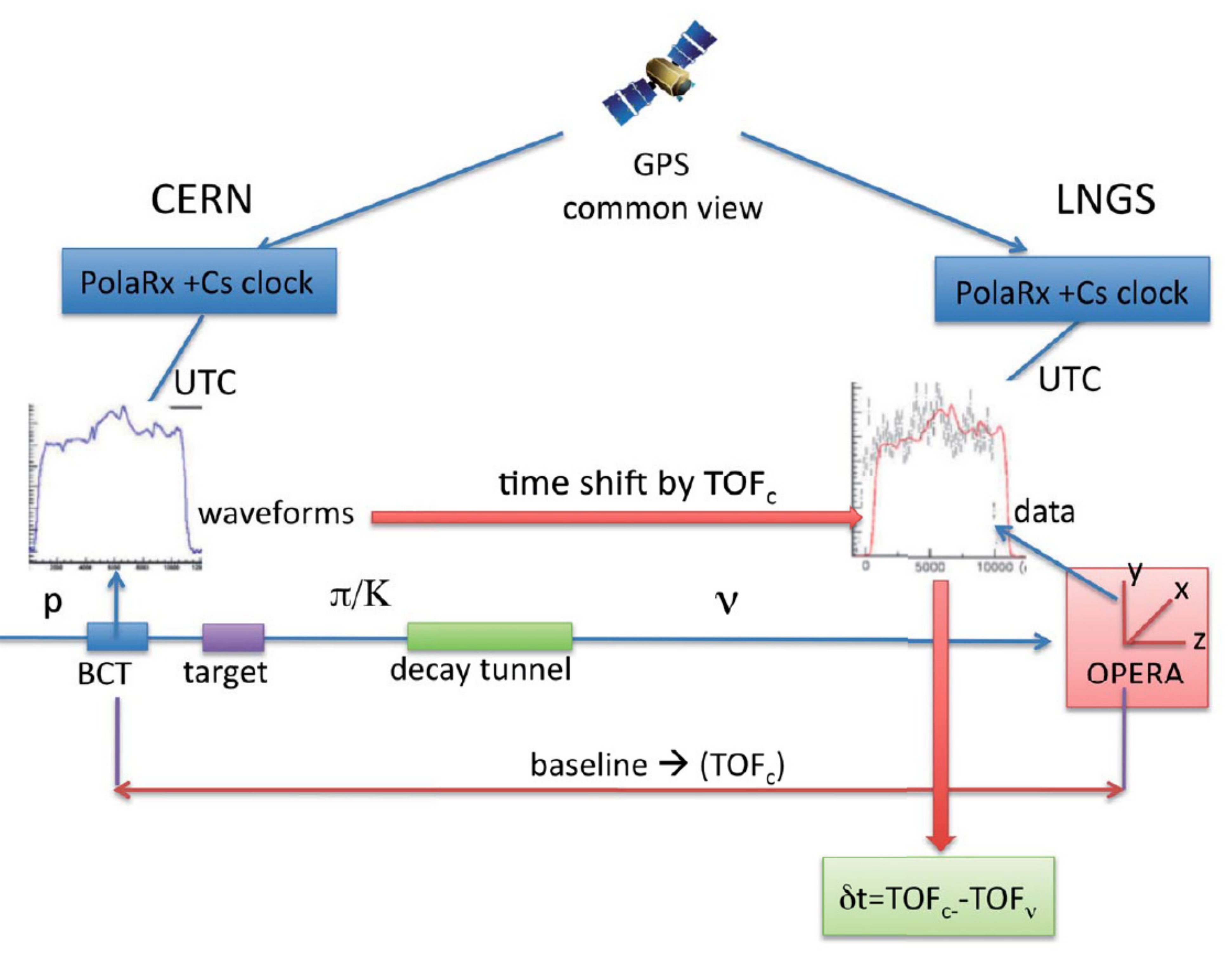}}}
\caption{Scheme of the time of flight measurement. Figure from \cite{6}.}
\label{strauss-fig15}
\end{myfigure}
The procedures are explained in great detail in \cite{6}. Since the time of the conference, an instrumental mistake has been identified that makes the results presented at this conference obsolete. Updated results taken from \cite{7b} are presented below.

\begin{myfigure}
\centerline{\resizebox{70mm}{!}{\includegraphics{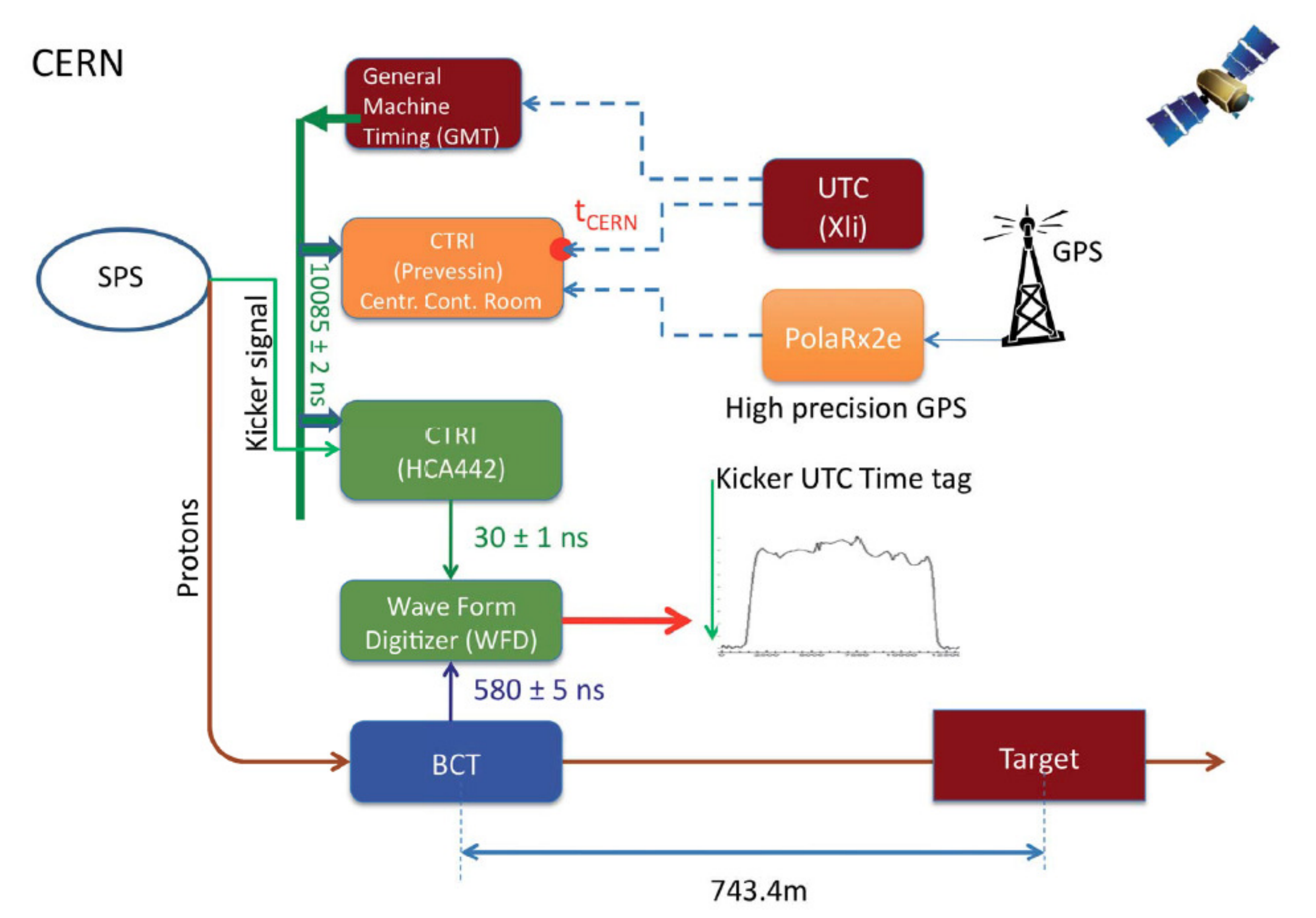}}}
\centerline{\resizebox{70mm}{!}{\includegraphics{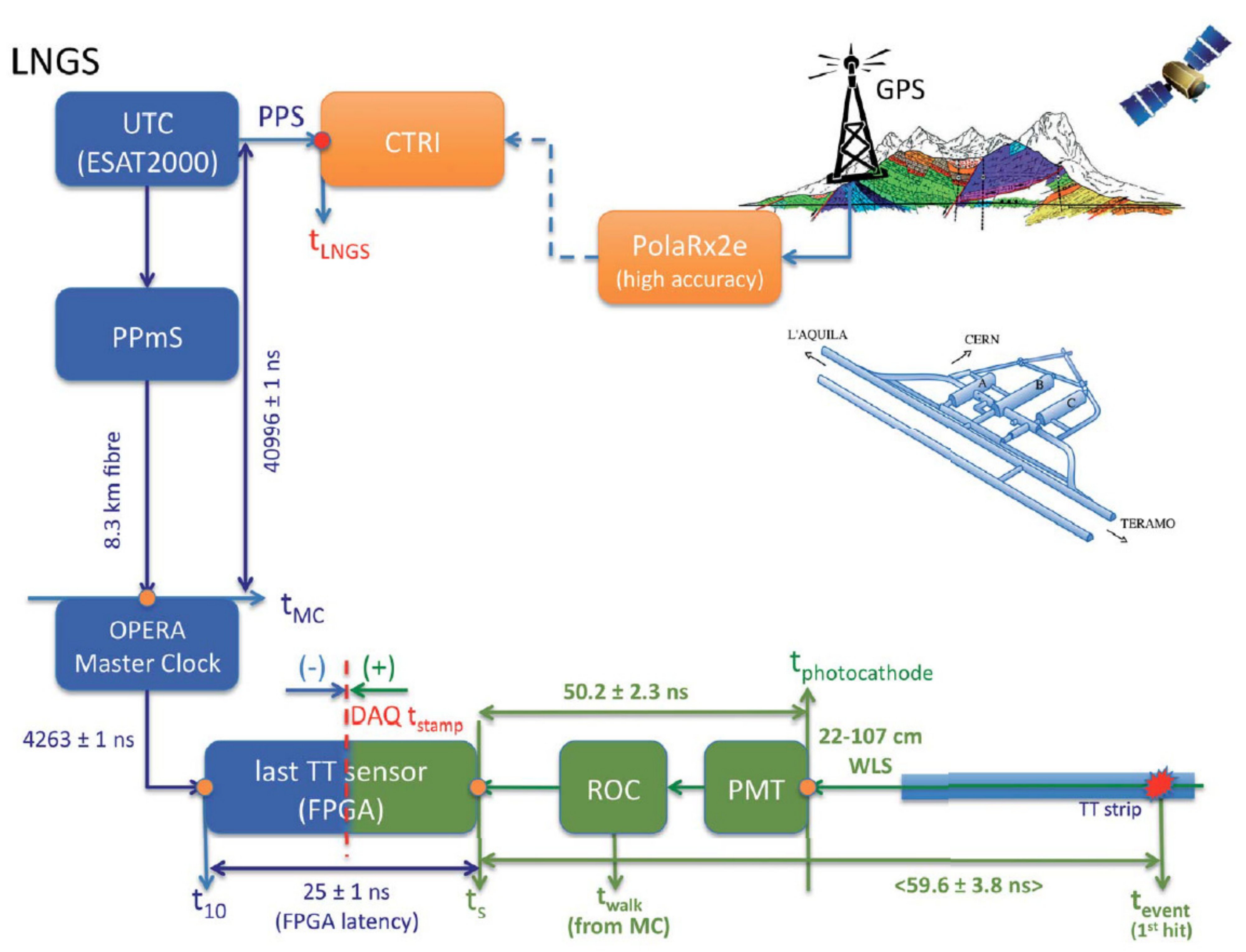}}}
\caption{Top: Scheme of the timing sytem at CERN. Bottom: Scheme of the timing system at LNGS. Figures from \cite{6}.}
\label{strauss-fig16}
\end{myfigure}

Fig. \ref{strauss-fig16} shows the timing systems both at CERN and LNGS, which allowed time calibration between both sites within accuracy of $\pm4$\,ns.

The distance between CERN and the OPERA detector was measured via GPS geodesy and extrapolation down to the location of both the CNGS target and the OPERA detector with terrestrial traverse methods. The effective baseline is measured as $731278.0\pm0.2$\,m.

The proton wave form for each SPS extraction was measured with a beam current transformer (BCT). The sum of the wave forms restricted to those associated to a neutrino interaction in OPERA was used as PDF for the time distribution of the events within the extraction. The maximum likelihood method was used to extract the time shift between the two distributions, i.e. the neutrino time of flight. Internal NC and CC interactions in the OPERA target and external CC interactions occurring in the upstream rock from the 2009, 2010 and 2011 CNG runs were used for this analysis. As shown in Fig. \ref{strauss-fig20}, it is measured to be: $$\delta t = \mbox{TOF}_c - \mbox{TOF}_\nu = (6.5\pm7.4(\mbox{stat.})\pm ^{+8.3}_{-8.0}(\mbox{syst.}))\,\mbox{ns.}$$ Modifying the analysis by using each neutrino interaction waveform as PDF instead of their sum gives a comparable result of
$$\delta t = (3.5\pm5.6(\mbox{stat.})\pm ^{+9.4}_{-9.1}(\mbox{syst.}))\,\mbox{ns.}$$ No energy dependence was observed.

\begin{myfigure}
\centerline{\resizebox{70mm}{!}{\includegraphics{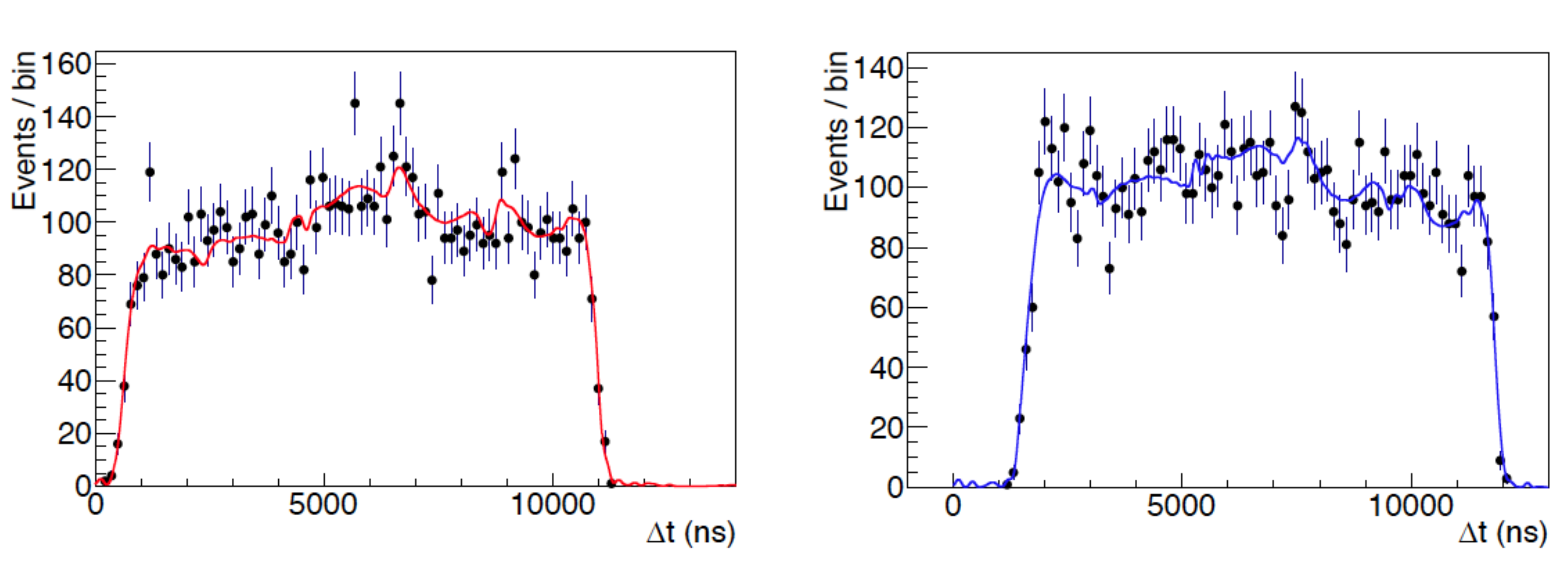}}}
\centerline{\resizebox{70mm}{!}{\includegraphics{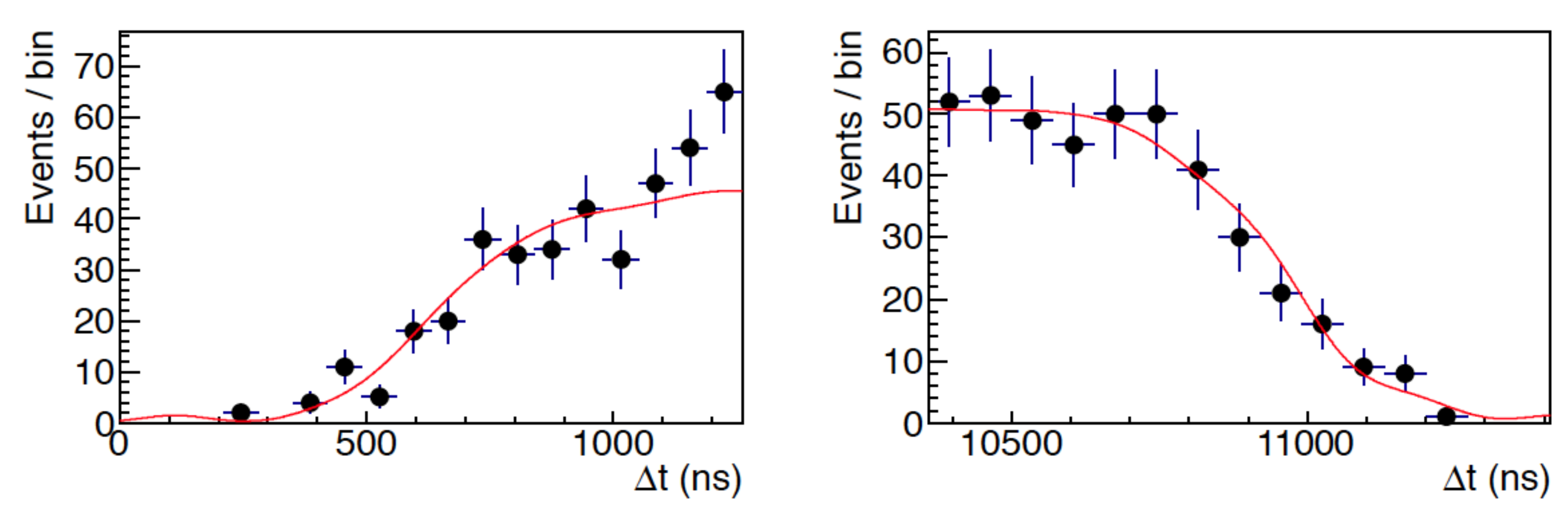}}}
\caption{Top: Comparison of the measured neutrino interaction time distributions (data points) and the proton PDF (red and blue line) for the two SPS extractions resulting from the maximum likelihood analysis. Bottom: Blow-up of the leading edge (left plot) and the trailing edge (right plot) of the measured neutrino interaction time distributions (data points) and the proton PDF (red line) for the first SPS extraction after correcting for $\delta t=6.5$\,ns. Within errors, this second extraction is equal to the first one. Figures from \cite{6}.}
\label{strauss-fig20}
\end{myfigure}

To cross-check for systematic effects, a dedicated bunched beam run was performed, where the SPS proton delivery is split into 3\,ns long spills, separated by 524\,ns in time during autumn 2011, and a similar mode of 3\,ns with 100\,ns separation in spring 2012. 

The value of $\delta t$ obtained in 2011 by using timing information provided by the target tracker is $1.9\pm3.7$\,ns; it is $0.8\pm3.5$\,ns when based on the spectrometer data \cite{6}. For the 2012 run, the corresponding values of $\delta t$ are $\delta t =(-1.6\pm1.1(stat.)^{+6.1}_{-3.7})$\,ns \cite{7b}. All results are in agreement with the measurement from standard CNGS beam operation.

\section{Conclusions}
The OPERA experiment detected two $\tau$ neutrino events appearing in the CNGS beam. Further, we measured the neutrino velocity to be in agreement with the speed of light in a vacuum to the O($10^{-6}$). Other short decay topologies like $\nu_e$ or charm decays can also be detected and are in agreement with MC expectations, thus providing a benchmark for validating the $\tau$ efficiency expectations.

\thanks
Firstly, I thank the organizers of the workshop and the OPERA PTB for the possibility to join this workshop. The OPERA collaboration thanks CERN, INFN and LNGS for their support and work. In addition OPERA is grateful for funding from the following national agencies: 

Fonds de la Recherche Scientique - FNRS and Institut Interuniversitaire des Sciences Nucl\'eaires for Belgium; MoSES for Croatia; CNRS and IN2P3 for France; BMBF for Germany; INFN for Italy; JSPS (Japan Society for the Promotion of Science), MEXT (Ministry of Education, Culture, Sports, Science and Technology), QFPU (Global COE program of Nagoya University, ``Quest for Fundamental Principles in the Universe", supported by JSPS and MEXT) and Promotion and Mutual Aid Corporation for Private Schools of Japan for Japan; The Swiss National Science Foundation (SNF), the University of Bern and ETH Zurich for Switzerland; the Russian Foundation for Basic Research (grant 09-02-00300 a), the Programs of the Presidium of the Russian Academy of Sciences ``Neutrino Physics" and ``Experimental and theoretical researches of fundamental interactions connected with work on the accelerator of CERN", the support programs of leading schools (grant 3517.2010.2), and the Ministry of Education and Science of the Russian Federation for Russia; the Korea Research Foundation Grant (KRF-2008-313-C00201) for Korea; and TUBITAK ``Scientific and Technological Research Council of Turkey", for Turkey. In addition the OPERA collaboration thanks the technical collaborators and the IN2P3 Computing Centre (CC-IN2P3).

\bigskip
\bigskip
\bigskip
\bigskip

\bigskip
\bigskip
\noindent {\bf DISCUSSION}

\bigskip
\noindent {\bf JAMES BEALL's Comment:} Do you have evidence that the neutrinos travel at less than the speed of light? When will we know the answer to this question?

\bigskip
\noindent {\bf THOMAS STRAUSS:} The speed of the neutrinos from CNGS to LNGS could be in agreement with the speed of light. The answer to the second question will be presented at the Neutrino 2012 conference. \textit{Addendum: The final results from \cite{6,7b} are used in these proceedings, but were not used in the talk at Vulcano 2012.}

\bigskip
\noindent {\bf MAURICE H.P.M. VAN PUTTEN's Comment:} In your upcoming announcement on neutrino velocity, will you insist on consistency of the results from the 0.5\,$\mu$s nuch experiment and the 2\,ns pulse experiment? In arxiv.org1110:4781,  I pointed out the need for a 2-parameter analysis for causal matched filtering, to accuarately determine the TOF. The results show a reduction to 3.75\,$\sigma$ from the OPERA claim of 6.04\,$\sigma$, demonstrating the need for a careful analysis. Will OPERA make its data public, and will OPERA pursue proper casual matching filtering on the 10.5\,$\mu$s experiment? 

\bigskip
\noindent {\bf THOMAS STRAUSS:} For the first question, the answer will be given at the Neutrino 2012 conference. The data is available to the public in the RAW format, and a guideline for setting up the timelink calibration has been developed together with CERN. As for the matched filtering, this question will be forwarded to the person responsible for the neutrino velocity analysis. \textit{Addendum (Statement from the Collaboration): The 2-parameter analysis has been debated at length. It is clear that by introducing a new degree of freedom (the length of the bunch) one may obtain a better result, as Putten actually obtains. However this corresponds to a possible new unknown physical property of neutrinos. To be conservative about the measurement in September, OPERA has chosen to exclude any debate on the possible physical source of the result.}

\end{multicols}

\begin{thebibliography}{99}

\bibitem{1} OPERA Collaboration, R. Acquafredda et al., JINST 4 (2009) P04018.
\bibitem{3a} Ed. K. Elsener, The CERN Neutrino beam to Gran Sasso (Conceptual Technical Design), CERN 98-02, INFN/AE-98/05;
\bibitem{3b} R. Bailey et al., The CERN Neutrino beam to Gran Sasso (CNGS) (Addendum to CERN 98-02, INFN/AE-98/05), CERN-SL/99-034(DI), INFN/AE-99/05.
\bibitem{2a} A. Ereditato, K. Niwa and P. Strolin, The emulsion technique for short, medium and long baseline $\nu_{\mu} \rightarrow \nu_{\tau}$ oscillation experiments, 423, INFN-AE-97-06, DAPNU-97-07;
\bibitem{2b} OPERA collaboration, H. Shibuya et al., Letter of intent: the OPERA emulsion detector for a long-baseline neutrino-oscillation experiment, CERN-SPSC-97-24, LNGS-LOI-8-97;
\bibitem{2c} OPERA collaboration, M. Guler et al., An appearance experiment to search for $\nu_{\mu}\rightarrow\nu_{\tau}$ oscillations in the CNGS beam: experimental proposal, CERN-SPSC-2000-028, LNGS P25/2000;
\bibitem{2d} OPERA collaboration, M. Guler et al., Status Report on the OPERA experiment;\\
CERN/SPSC 2001-025, LNGS-EXP 30/2001 add. 1/01;
\bibitem{2e} OPERA Collaboration, N. Agafonova et al., Phys. Lett. B 691 (2010) 138;
\bibitem{2f} OPERA Collaboration, N. Agafonova et al., arXiv:1107.2594v1.
\bibitem{4} N. Agafonova, Search for nu-mu - nu-tau oscillation with the OPERA experiment in the CNGS beam, New J. Phys. 14 (2012) 033017.
\bibitem{7a} M. Nakamura, Neutrino 2012, XXV International Conference on Neutrino Physics and Astrophysics, 3-9 June 2012, Kyoto, Japan.
\bibitem{5}  Strauss, T., Charm production in the OPERA experiment and the study of a high temperature superconducting solenoid for a liquid argon time projection chamber, PhD thesis ``ETH-19247".
\bibitem{6} OPERA Collaboration, T. Adam et al., arXiv:1109.4897v4.
\bibitem{7b} M. Dracos, Neutrino 2012, XXV International Conference on Neutrino Physics and Astrophysics, 3-9 June 2012, Kyoto, Japan.
\end{thebibliography}
\end{document}